\documentclass[twocolumn,showkeys,amsmath,amssymb,prl]{revtex4-1}
\usepackage{epsf}
\usepackage{amsmath,amsthm,amssymb}
\usepackage{color}
\usepackage{dcolumn}
\usepackage{bm}
\usepackage{graphicx,epsfig}
\usepackage[dvipsnames]{xcolor}

\graphicspath{{fig/}}
\everymath{\displaystyle}

\begin{document} 

\title{General quantum-mechanical solution for twisted electrons in a uniform magnetic field}

\author{Liping Zou$^{1,3}$}
\email{zoulp@impcas.ac.cn}
\author{Pengming Zhang$^{2}$}
\email{corresponding author. zhangpm5@mail.sysu.edu.cn}
\author{Alexander J. Silenko$^{3,4,5}$}
\email{alsilenko@mail.ru}

\affiliation{$^1$Sino-French Institute of Nuclear Engineering and Technology, Sun Yat-Sen University, Zhuhai 519082, China}
\affiliation{$^2$School of Physics and Astronomy, Sun Yat-sen University, Zhuhai 519082, China}
\affiliation{$^3$Institute of Modern Physics, Chinese Academy of
Sciences, Lanzhou 730000, China}
\affiliation{$^4$Bogoliubov Laboratory of Theoretical Physics, Joint
Institute for Nuclear Research, Dubna 141980, Russia}
\affiliation{$^5$Research Institute for Nuclear Problems, Belarusian
State University, Minsk 220030, Belarus}

\date{\today}

\begin{abstract}
A theory of twisted (and other structured) paraxial electrons in a uniform magnetic field is developed. The obtained general quantum-mechanical solution of the relativistic paraxial equation contains the commonly accepted result as a specific case of unstructured electron waves. Unlike all precedent investigations, the present study describes structured electron states which are not plane waves along the magnetic field direction. In the weak-field limit, our solution (unlike the existing theory) is consistent with the well-known equation for free twisted electron beams. The observable effect of a different behavior of relativistic Laguerre-Gauss beams with opposite directions of the orbital angular momentum penetrating from the free space into a magnetic field is predicted. Distinguishing features of the quantization of the velocity and the effective mass of the Laguerre-Gauss and Landau electrons in the uniform magnetic field are analyzed.
\end{abstract}


\maketitle


The discovery of twisted (vortex) electron states with a nonzero intrinsic orbital angular momentum (OAM) \cite{UTV} has confirmed their theoretical prediction \cite{Bliokh2007}
and has created new applications of electron beams. Twisted electrons are successfully used in the electron microscopy and in investigations of magnetic phenomena
(see Refs. \cite{BliokhSOI,Lloyd,LarocqueTwEl,LloydPhysRevLett2012,Rusz,Edstrom,imaging,Observation,OriginDemonstration}
and references therein). Twisted electron beams with large intrinsic OAMs (up to 1000$\hbar$)
have been recently obtained \cite{VGRILLO}. Due to large magnetic moments of twisted electrons, their above-mentioned applications are very natural. This situation makes a correct and full description of twisted electrons in a magnetic field to be very important.

In the present study, we use the system of units $\hbar=1,~c=1$. We include $\hbar$ and $c$ explicitly when this inclusion
clarifies the problem.

Let us direct the $z$ axis of the cylindrical coordinates $r,\phi,z$ along the uniform magnetic field, $\bm B=B\bm e_z$. It is now generally accepted \cite{Bliokhmagnetic,magnetic,Kruining,Rajabi,Greenshields,Greenshields2015} that twisted electron states in a uniform magnetic field are defined by the Landau wave function \cite{Landau,LL3} or its relativistic generalizations \cite{Rajabi,Energy2a,Energy2,OConnell,CanutoChiu}. This function being an eigenfunction of the nonrelativistic Hamiltonian
\begin{equation}
{\cal H}=\frac{\bm{\pi}^2-e\bm\sigma\cdot\bm B}{2m},\quad \bm{\pi}^2=-\nabla^2+ieB\frac{\partial}{\partial\phi}+\frac{e^2B^2r^2}{4}
\label{HL}
\end{equation}
reads
\begin{equation}
\begin{array}{c}
\psi={\cal A}\exp{(i\ell\phi)}\exp{(ip_zz)},\qquad \int{\psi^\dag\psi rdrd\phi}=1,\\
{\cal A}=\frac{C_{n\ell}}{w_m}\left(\frac{\sqrt2r}{w_m}\right)^{|\ell|}
L_n^{|\ell|}\left(\frac{2r^2}{w^2_m}\right)\exp{\left(-\frac{r^2}{w^2_m}\right)}\eta,\\
C_{n\ell}=\sqrt{\frac{2n!}{\pi(n+|\ell|)!}},\qquad
w_m=\frac{2}{\sqrt{|e|B}}.
\end{array}
\label{Lenergy}
\end{equation}
Here $\bm{\pi}=\bm{p}-e\bm A$ is the kinetic momentum, the real function ${\cal A}$ defines the beam amplitude, $L_n^
{|\ell|}$ is the generalized Laguerre polynomial, and $n = 0, 1, 2,\dots$ is the radial quantum number. When the cylindrical coordinates are used, $A_\phi=Br/2,\,A_r=A_z=0$. For the electron, $e=-|e|$. The spin function $\eta$ is an eigenfunction of the Pauli operator $\sigma_z$ (cf. Ref. \cite{Energy3}):
\begin{equation}\sigma_z\eta^\pm=\pm\eta^\pm,\quad
\eta^+=\left(\begin{array}{c} 1 \\ 0 \end{array}\right),\quad \eta^-=\left(\begin{array}{c} 0 \\ 1 \end{array}\right).
\label{spineta}
\end{equation}
The distinctive feature of the Landau solution is the trivial (exponential) dependence of the electron wave function on $z$. Values of $p_z$ are fixed and $\psi$ is an eigenfunction of the operator $p_z\equiv -i\hbar\partial/(\partial z)$.

The twisted states of free photons and electrons are defined by the paraxial wave equation \cite{BliokhSOI,BBP,Siegman}:
\begin{equation}
\begin{array}{c}
\left(\nabla^2_\bot+2ik\frac{\partial}{\partial
z}\right)\!\Psi=0,\quad
\nabla^2_\bot=
\frac{\partial^2}{\partial r^2}+\frac1r\frac{\partial}{\partial
r}+\frac{1}{r^2}\frac{\partial^2}{\partial\phi^2}.
\end{array}
\label{eqp}
\end{equation} For electrons, it can be obtained from the Dirac equation in the Foldy-Wouthuysen (FW) representation provided that $|\bm p_\bot|\ll p$ \cite{photonPRA}. The paraxial wave function of free electrons and photons characterizes the Laguerre-Gauss (LG) beams and reads \cite{BBP,Allen,PlickKrenn}
\begin{equation}
\begin{array}{c}
\Psi=\mathbb{A}\exp{(i\Phi)},\qquad \int{\Psi^\dag\Psi rdrd\phi}=1,\\
\mathbb{A}=\frac{C_{n\ell}}{w(z)}\left(\frac{\sqrt2r}{w(z)}\right)^{|\ell|}
L_n^{|\ell|}\left(\frac{2r^2}{w^2(z)}\right)\exp{\left(-\frac{r^2}{w^2(z)}\right)}\eta,\\
\Phi=\ell\phi+\frac{kr^2}{2R(z)}-\Phi_G(z),
\end{array}
\label{eq33new}
\end{equation} where
\begin{equation}
\begin{array}{c}
w(z)=w_0\sqrt{1+\frac{z^2}{z_R^2}},\quad
R(z)=z+\frac{z_R^2}{z},\quad
z_R=\frac{kw_0^2}{2},\\
\Phi_G(z)=N\arctan{\left(\frac{z}{z_R}\right)},\quad N=2n+|\ell|+1,
\end{array}
\label{eqaddit}
\end{equation}
the real functions $\mathbb{A}$ and $\Phi$ define the amplitude and phase,
$k$ is the beam wavenumber, $w_0$ is the beam
waist (minimum beam width), $R(z)$ is the radius of curvature of the wavefront, $\Phi_G(z)$ is the Gouy phase, and $z_R$ is the Rayleigh diffraction length. The quantities $C_{n\ell}$ and $\eta$ are given by Eqs. (\ref{Lenergy}) and (\ref{spineta}), respectively. For electrons, $\Psi$ is a spinor. Evidently, $\Psi$ is not an eigenfunction of the operator $p_z$. Therefore, the free-space wave function (\ref{eq33new}), (\ref{eqaddit}) characterizes a beam formed by partial waves with
different $p_z$.

A correspondence between the relativistic quantum-mechanical equations in the FW representation and the paraxial wave equations has been established in Refs. \cite{photonPRA,Barnett-FWQM}. The correspondence is very similar for photons and electrons. In connection with this similarity, we can mention the existence of bosonic symmetries of the standard Dirac equation \cite{Simulik,Simulik1,Simulik2,Simulik3,Simulik4,Simulik5,Simuliknew}.

Advanced results obtained in optics allow us to rigorously derive a general formula for the paraxial wave function of a relativistic twisted Dirac particle in a uniform magnetic field. In this case, the exact relativistic FW Hamiltonian is given by
\cite{Energy3,Case,Energy1,JMP}
\begin{equation}
\begin{array}{c}
i\frac{\partial\Psi_{FW}}{\partial t}=\!{\cal H}_{FW}\Psi_{FW},\quad {\cal H}_{FW}=\!\beta\sqrt{m^2+\bm{\pi}^2-e\bm\Sigma\cdot\bm B},
\end{array}
\label{eq12new}
\end{equation}
where $\bm{\pi}=\bm{p}-e\bm A$ is the kinetic momentum and $\beta$ and $\bm\Sigma$ are the Dirac matrices.
This Hamiltonian acts on the bispinor $\Psi_{FW}=
\left(\begin{array}{c} \Phi_{FW} \\ 0 \end{array}\right)$. The zero lower spinor of the bispinor can be disregarded.
Eigenfunctions (more precisely, an upper spinor) of the \emph{relativistic} FW Hamiltonian coincide with the \emph{nonrelativistic} Landau solution (\ref{Lenergy}) because the operator $\bm{\pi}^2-e\bm\Sigma\cdot\bm B$ commutes with the Hamiltonian in both cases (see Refs. \cite{Energy3,Case,Energy1}). The FW representation is important for obtaining a classical limit of relativistic quantum mechanics \cite{JINRLett12} and establishing a connection between relativistic and nonrelativistic quantum mechanics \cite{Reply2019,PRAFW}.

Let us denote $P=\sqrt{E^2-m^2}=\hbar k$, where $E$ is an energy of a stationary state. A transformation of Hamiltonian equations in the FW representation to the paraxial form has been considered in Refs. \cite{photonPRA,Barnett-FWQM,paraxialLandau}. Squaring
Eq. (\ref{eq12new}) for the upper spinor, 
applying the paraxial approximation for $p_z>0$, 
and the substitution $\Phi_{FW}=\exp{(ikz)}\Psi$ lead to the 
paraxial equation \cite{paraxialLandau}
\begin{equation}
\begin{array}{c}
\left(\nabla^2_\bot-ieB\frac{\partial}{\partial\phi}-\frac{e^2B^2r^2}{4}+2es_zB+2ik\frac{\partial}{\partial
z}\right)\Psi=0, 
\end{array}
\label{eqpar}
\end{equation} where $s_z$ is the spin
projection onto the field direction. The above-mentioned substitution is equivalent to shifts of the zero energy level and of the squared particle momentum in Schr\"{o}dinger quantum mechanics.
When $B=0$, Eq. (\ref{eqpar}) takes the form of the paraxial wave equation for free electrons (\ref{eqp}).

The paraxial form of the Landau wave function is an eigenfunction of Eq. (\ref{eqpar}) and is given by \cite{paraxialLandau}
\begin{equation}
\begin{array}{c}
\Psi={\cal A}\exp{(i\ell\phi)}\exp{[-i\zeta_G(z)]},\\
\zeta_G(z)=\left(2n+1+|\ell|+\ell+2s_z\right)\frac{2z}{kw_m^2},\quad w_m=\frac{2}{\sqrt{|e|B}},
\end{array}
\label{Lparaxl}
\end{equation} where $\zeta_G(z)$ is the Gouy phase. Amazingly, the probability and charge densities defined by the \emph{nonrelativistic} Landau wave function (\ref{Lenergy}) and by the \emph{relativistic} paraxial wave function (\ref{Lparaxl}) coincide. This property follows from the paraxial approximation for the electron velocity, $|v_\perp|\ll v$, and takes place in the laboratory frame. In this frame, the transversal motion can be described by means of nonrelativistic quantum mechanics.

While Eq. (\ref{Lparaxl}) is similar to Eqs. (\ref{eq33new}) and (\ref{eqaddit}), there is a substantial
difference between them. The paraxial Landau wave function (\ref{Lparaxl}) describes a wave with a fixed value of
$p_z$, while the free-space LG beam is formed by partial waves with
different $p_z$. Thus, the use of the function (\ref{Lparaxl}) for a \emph{general} description of a twisted paraxial
electron in a uniform magnetic field means that even a weak magnetic field leads to destroying the longitudinal
structure of LG beams. However, any partial wave forming the beams does not change its energy during the beam penetration from the free space into a magnetic field region. Therefore, the above-mentioned meaning is not reasonable and the Landau solution of Eq. (\ref{eqpar}) is not
general. To obtain the general solution of this equation, we use its similarity to Eq. (\ref{eqp}) and utilize optical
approach \cite{Siegman,Kogelnik,Alda,Pampaloni} applied for
the free-space paraxial equation (see Supplemental Material
\cite{SupplementalMaterial}, Sec. I). 
The subsequent derivation shows that an electron state is described by the matter wave beam (\ref{eq33new}) but three functions on $z$ should be overridden. The power and exponential functions are defined by an asymptotic behavior of Eq. (\ref{eqpar}) at $\zeta\rightarrow0$ and $\zeta\rightarrow\infty$, respectively (see Ref. \cite{LL3}).
These functions cannot be totally specified \emph{a priori} because of the presence of the last operator in Eq.  (\ref{eqpar}). It is helpful to suppose that the general solution of Eq. (\ref{eqpar}) has the form (\ref{eq33new}), where $w(z),\,R(z)$, and $\Phi_G(z)$ are not yet specified. First of all, we need to check that the corresponding wave functions $\Psi$ form a set of orthogonal eigenfunctions. When we denote $$\zeta=\frac{2r^2}{w^2(z)},$$
%
the substitution of $\Psi$ into Eq. (\ref{eqpar}) results in
\begin{equation}
\begin{array}{c}
\exp{[i(\Phi-\ell\phi)]}\nabla^2_\bot\Psi_0+\Biggl\{-\frac{4(\ell+2s_z)}{w_m^2}-\frac{4r^2}{w_m^4}\\-\frac{k^2r^2}{R^2(z)}-k^2r^2\biggl[\frac{1}{R(z)}\biggr]'+2k\Phi'_G(z)\Biggr\}\Psi \\
+2ik\Upsilon(z)\left[1+|\ell|-\frac{2r^2}{w^2(z)}+\frac{4r^2{L_n^{|\ell|}}'(\zeta)}{w^2(z)L_n^{|\ell|}(\zeta)}\right]\Psi=0, \\ \nabla^2_\bot\Psi_0=\frac{2}{w^2(z)}\biggl[4\zeta {L_n^{|\ell|}}''(\zeta)+ 4(-\zeta+|\ell|+1){L_n^{|\ell|}}'(\zeta)\\+(\zeta-2|\ell|-2)L_n^{|\ell|}(\zeta)\biggr]\frac{\Psi_0}{L_n^{|\ell|}(\zeta)},\\
\Upsilon(z)=\frac{1}{R(z)}-\frac{w'(z)}{w(z)},
\end{array}
\label{eqparnw}
\end{equation} where $\Psi_0=\mathbb{A}\exp{(i\ell\phi)}$ and primes denote derivatives with respect to mentioned variables ($\zeta$ or $z$). 
We can check that properties of the generalized Laguerre polynomials confirm our supposition about the validity of wave functions (\ref{eq33new}) in the considered case. In this case,
\begin{equation}
\begin{array}{c}
\nabla^2_\bot\Psi_0=\frac{4}{w^2(z)}\biggl[\frac{r^2}{w^2(z)}-N\biggr]\Psi_0,
\end{array}
\label{eqLanw}
\end{equation} $\Psi_0$ coincides with the Landau wave eigenfunction,
and the following conditions should be satisfied:
\begin{equation}
\begin{array}{c}
\frac{1}{R(z)}=\frac{w'(z)}{w(z)}, \quad\frac{k^2}{R^2(z)}+k^2\biggl[\frac{1}{R(z)}\biggr]'=\frac{4}{w^4(z)}-\frac{4}{w_m^4}, \\ 2k\Phi'_G(z)=\frac{4(\ell+2s_z)}{w_m^2}+\frac{4N}{w^2(z)}.
\end{array}
\label{condt}
\end{equation}

The straightforward solution of these differential equations is based on known integrals \cite{GradshteynRyzhik} and has the form (Supplemental Material
\cite{SupplementalMaterial}, Sec. II)
\begin{equation}
\begin{array}{c}
w(z)=w_0\sqrt{\frac12\left[1+\frac{w_m^4}{w_0^4}-\left(\frac{w_m^4}{w_0^4}-1\right)\cos{\frac{2z}{z_m}}\right]}\\=w_0\sqrt{\cos^2{\frac{z}{z_m}}+\frac{w_m^4}{w_0^4}\sin^2{\frac{z}{z_m}}},\qquad z_m=\frac{kw_m^2}{2},\\
R(z)=kw_m^2\frac{\cos^2{\frac{z}{z_m}}+\frac{w_m^4}{w_0^4}\sin^2{\frac{z}{z_m}}}{\left(\frac{w_m^4}{w_0^4}-1\right)\sin{\frac{2z}{z_m}}},\\ \Phi_G(z)=N\arctan{\left(\frac{w_m^2}{w_0^2}\tan{\frac{z}{z_m}}\right)}+\frac{(\ell+2s_z)z}{z_m}.
\end{array}
\label{general}
\end{equation} The normalization constant $C_{n\ell}$ is given by Eq. (\ref{Lenergy}).

It is important that the exact FW Hamiltonian for a Dirac particle in a \emph{nonuniform} but time-independent magnetic field $\bm B(\bm r)$ has also the form (\ref{eq12new}) \cite{Case,JMP}. We expect that our approach can be useful for a general description of relativistic Dirac particle beams in some axially symmetric nonuniform magnetic fields, in particular, for the relativistic electron beams in round magnetic lenses and real solenoids. These problems are of great practical importance (see Refs. \cite{Loffler,FloettmannKarlovets} and references therein).

\begin{center}
\begin{figure}
\includegraphics[width=0.35\textwidth]{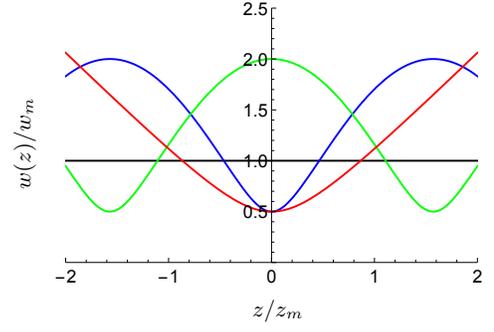}
\caption{The beam width, $w(z)$, of a twisted electron beam in a uniform magnetic field for different values of the ratio $w_0/w_m$. The black line describes the case of $w(z)=w_0=w_m$, when the beam width is equal to the transverse magnetic width of Landau levels, $w_m$. The blue and green lines demonstrate the beam width defined by our general solution for $w_0=0.5w_m$ and $w_0=2 w_m$, respectively. The red line shows the beam width of a \emph{free} twisted electron beam for $w_0=0.5 w_m$.}
\label{fig1}
\end{figure}
\end{center}

\begin{center}
\begin{figure}
\includegraphics[width=0.475\textwidth]{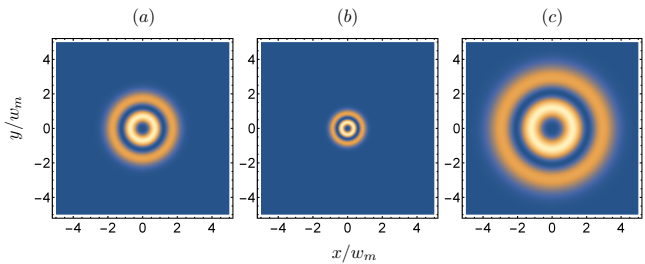}
\caption{The transverse probability density defined by our solution for different values of the longitudinal coordinate $z$ and the ratio $w_0/w_m$. In the case (a), $w_0=w_m$ and our general solution coincides with the Landau solution and is independent of the longitudinal position. In the case (b), $w_0=0.5w_m$ and $z=0$. In this case, our solution coincides with the corresponding one for a free twisted electron. In the case (c), $w_0=0.5w_m$ and $z=z_m$. All the plots presented correspond to the quantum numbers $n=1,~\ell=2$.}
\label{fig2}
\end{figure}
\end{center}

In the cases of $w_m>w_0$ and $w_m<w_0$ (for a relatively weak and strong magnetic field, respectively), the derivations are very different but the corresponding formulas for $w(z),R(z)$, and $\Phi_G(z)$ coincide. $w(z)$ oscillates between $w_0$ and $w_m^2/w_0$. In the case of $B\rightarrow0$ ($w_m>>w_0$), $z<<z_m$, there is a full compliance with the solution for a free twisted particle and the beam parameters (\ref{general}) take the form (\ref{eqaddit}). This important property is illustrated by Fig. \ref{fig1} (the blue and red lines at $z=0$) and Fig. \ref{fig2} (the middle plot). In the latter figure, the probability density distribution in the $xy$ plane is shown.  Our result coincides with the Landau solution when $w_0=w_m$. The coincidence is shown by the black line at Fig. \ref{fig1} and by Fig. \ref{fig2} (the left plot). In this case, the general wave function (\ref{eq33new}), (\ref{general}) takes the form (\ref{Lparaxl}) and $w(z)=w_m=const$. However, 
the paraxial form (\ref{Lparaxl}) of the Landau wave eigenfunction cannot 
explain a transition to the free-space solution (\ref{eq33new}), (\ref{eqaddit}) at $B\rightarrow0$.
The inconsistency of the weak-field limits of the Landau wave eigenfunction and its relativistic generalizations with the well-known equation for free twisted electron beams is rather natural because Refs. \cite{Landau,LL3,Rajabi,Energy2a,Energy2,OConnell,CanutoChiu} describe only unstructured electrons.

Unlike the wave function in the free space, the wave function defined by Eq. (\ref{general}) is spatially periodic. Amazingly, its period $\mathfrak{L}=\pi z_m=\pi kw_m^2/2=2\pi P/(\omega_cE)$ is equal to the pitch of the helix characterizing the classical motion of electrons ($\omega_c$ is the cyclotron frequency). 
The spatially periodic behavior of the wave function is illustrated by Fig. \ref{fig1}. A similarity between the probability density distributions of the LG and Landau wave eigenfunctions is demonstrated by Fig. \ref{fig2}.
Equations (\ref{eq33new}) and (\ref{general}) show that the wave function of a twisted electron in a uniform magnetic field depends only on the total OAM $\hbar\ell$ but not on intrinsic and extrinsic OAMs separately. Therefore, the two latter OAMs cannot be separated.

The LG beam described by Eqs. (\ref{eq33new}), (\ref{general}) is formed by partial waves with the same $E$ and $P$ but slightly different directions of the kinetic momentum.
Figures \ref{fig1}, \ref{fig2} clearly show differences between the Landau solution and our one. As follows from Eq. (\ref{general}) and Fig. \ref{fig1}, our solution for $w(z)$ always crosses the Landau line.

The mean square of the beam radius can be obtained by an integration of the operator $r^2$ over the \emph{transversal} coordinates $r,\phi$ and reads (cf. Refs. \cite{Bliokhmagnetic,PhysRevLettLanzhou2019})
\begin{equation}
<r^2>=\int{\Psi^\dag\Psi r^3drd\phi}=\frac{w^2(z)}{2}\left(2n+|\ell|+1\right).
\label{eqrquar} \end{equation}
The electric quadrupole moment of twisted electrons introduced in Ref. \cite{PhysRevLettLanzhou2019} is measured in the focal plane $z=0$ and is given by
\begin{equation}
Q_0=\frac{|e|w^2_0}{2}\left(2n+|\ell|+1\right).
\label{eqquadr} \end{equation} The relativistic magnetic moment and the tensor magnetic polarizability are defined in Refs. \cite{Barut} and \cite{PhysRevLettLanzhou2019}, respectively (see also Refs. \cite{ResonanceTwistedElectrons,NPCS2018}).
We note that integrating over the \emph{longitudinal} coordinate results in
\begin{equation}
\begin{array}{c}
\frac{1}{2\pi z_m}\int_0^{2\pi z_m}{w^2(z)dz}=\frac{w_0^2}{2}\left(1+\frac{w_m^4}{w_0^4}\right)\\=w_m^2\left[1+\frac{1}{2}\left(\frac{w_m}{w_0}-\frac{w_0}{w_m}\right)^2\right].
\end{array}
\label{note}
\end{equation}

Equation (\ref{Lenergy}), its relativistic generalizations \cite{Energy2a,Energy2,Rajabi}, and Eq. (\ref{Lparaxl}) do not describe twisted electrons which constitute \emph{structured} beams even in the free space. The necessity to use the general equations (\ref{eq33new}), (\ref{general}) should substantially change the present theoretical description \cite{BliokhSOI,Rusz,Edstrom,Bliokhmagnetic,magnetic,Kruining} of twisted electron beams in uniform magnetic fields.

Energies of all partial waves which manifold defines a twisted or a untwisted structured state conserve when a beam penetrates from the free space into a magnetic field region. Therefore, final energies of such partial waves also coincide. This property remains valid for any nonuniform magnetic field. We predict the effect of a different behavior of two LG beams with \emph{opposite} OAM directions penetrating from the free space into the magnetic field. Due to a helical motion in the magnetic field, both twisted and untwisted electrons acquire \emph{extrinsic} OAMs with \emph{positive} projections onto the field direction (cf. Ref. \cite{paraxialLandau}). When the initial \emph{intrinsic} OAMs of electrons in the two beams are antiparallel ($\ell_1=-\ell_2=|\ell_2|$), an appearance of the \emph{extrinsic} OAMs conditions the relation $\ell'_1+\ell'_2>0$ for the final OAMs. The change of total OAMs leads to the difference of magnetic moments of electrons which is observable.
The transversal velocities of twisted electrons are nonzero and the Lorentz force turns electrons inwards and outwards for the beams with the intrinsic OAMs $\ell_1>0$ and $\ell_2<0$, respectively. When the initial beam waists coincide ($w_1=w_2$), the final beam waists differ and $w_1'<w_2'$. The difference should be of the order of $w_m$. In the general case, the radial quantum numbers will also be changed ($n_1'\neq n_2'$). The effect is observable and the predicted properties can be discovered in a \emph{specially designed} experiment which is in principle similar to that fulfilled in Ref. \cite{Schattschneider}.

In Ref. \cite{photonPRA}, the effect of a quantization of the velocity and the effective mass of structured photons and electrons has been predicted. A similar effect should takes place for twisted and other structured electrons in the uniform magnetic field. Expectation values of the group velocity are obtained by integrating the operator $\bm v=\partial{\cal H}_{FW}/(\partial\bm p)$ over the \emph{transversal} coordinates $r,\phi$ and are defined by (cf. Ref. \cite{photonPRA})
\begin{equation}
\begin{array}{c}
<v_z>=\frac{cP}{E}\left(1-\frac{<\bm\pi^2_\bot>-2es_zB}{2P^2}\right)
\\=\frac{ck}{\sqrt{k^2+K^2}}\left(1+\frac{1}{k}\left\langle\frac{\partial\Phi}{\partial z}\right\rangle\right),\qquad K=\frac{mc}{\hbar}.
\end{array}
\label{veloz} \end{equation} The effective electron mass is equal to (cf. Refs. \cite{photonPRA,Karlovets})
\begin{equation}
\begin{array}{c}
m_{eff}=\sqrt{m^2+<\bm\pi_\bot^2>-2es_zB}=\sqrt{m^2-2k\left\langle\frac{\partial\Phi}{\partial z}\right\rangle}.
\end{array}
\label{massz} \end{equation}
Cumbersome but straightforward calculations similar to those fulfilled in Ref. \cite{photonPRA} result in
\begin{equation}
\begin{array}{c}
<v_z>=\frac{ck}{\sqrt{k^2+K^2}}\biggl[1-\frac{\Lambda}{k}\biggr],\quad
m_{eff}=\sqrt{m^2+2k\Lambda},\\ \Lambda=-\left\langle\frac{\partial\Phi}{\partial z}\right\rangle=
\frac{N}{k}\left(\frac{1}{w_0^2}+\frac{w_0^2}{w_m^4}\right)+\frac{2(\ell+2s_z)}{kw_m^2}.
\end{array}
\label{eqphfel}
\end{equation}

In the uniform magnetic field, the effect of quantization of the velocity and the effective mass of structured electrons strongly depends on a kind of the structured electron beam.
A distance between neighboring quantum levels of these quantities defining the quantization is determined by $\Lambda$. Importantly, a similar quantization takes place for the Landau beams in the uniform magnetic field and the LG beams in the free space. In these cases, the formulas for $<v_z>$ and $m_{eff}$ presented in Eq. (\ref{eqphfel}) remain the same
and the corresponding formulas for $\Lambda_L$ and $\Lambda_{free}$ read
\begin{equation}
\begin{array}{c}
\Lambda_L=\frac{2(N+\ell+2s_z)}{kw_m^2},\qquad \Lambda_{free}=\frac{N}{kw_0^2}.
\end{array}
\label{eqlambda}
\end{equation} The formula for $\Lambda_{L}$ follows from the paraxial form of the Landau wave function (\ref{Lparaxl}) and can also be extracted from the Landau solution.
The formula for $\Lambda_{free}$ has been derived in Ref. \cite{photonPRA}. These formulas can also be obtained from Eq. (\ref{eqphfel}). For nonoscillating Landau beams $w_0=w_m$ and for twisted beams in the free space $w_m\rightarrow\infty~(B=0)$. While some enhancement of $\Lambda$ takes place in comparison with $\Lambda_{free}$ due to the presence of the uniform magnetic field, this enhancement is not significant.

The kinetic energy of twisted electrons usually obtained in electron microscopes is about 200$\div$300 keV. An estimate of the beam waist $w_0\sim10^{-9}$ m can be extracted from results obtained in Ref. \cite{OAMdiameter}. When $B=1$ T, $w_m=5.1\times10^{-8}$ m. In most cases $w_m$ is, therefore, significantly bigger than $w_0$. Equations (\ref{eqphfel}) and (\ref{eqlambda}) explicitly shows that the distances between neighboring quantum levels of average velocities of the Landau and LG beams in magnetic fields also substantially differ. As follows from Eq. (\ref{veloz}), the average velocities themselves coincide for the two kinds of beams if the wavenumber $k$ is the same. For the LG beams, the above-mentioned distance is defined by
\begin{equation}
\begin{array}{c}
\Delta<v_z>=\frac{\Delta<v_z>_{free}\Lambda}{\Lambda_{free}}\\=\frac{c}{k\sqrt{k^2+K^2}w_0^2}\left[1+\frac{w_0^4}{w_m^4}+\frac{2(\ell+2s_z)w_0^2}{Nw_m^2}\right].
\end{array}
\label{eqlambd}
\end{equation}
As a rule, $\Lambda$ is close to $\Lambda_{free}$. For the above-mentioned $w_0$ and $w_m$, $s_z=1/2$, and the quantum state illustrated by Fig. \ref{fig2}, $\Lambda/\Lambda_{free}=1.0005$.
When the kinetic energy $E-mc^2=200$ keV and $w_0=10^{-9}$ m, $\Delta<v_z>= 1.1\times10^{-7} c= 33$ m/s. When the beam energy spread is small enough, this quantity may be measured. Thus, the existence of the LG beams in the uniform magnetic field can be proven. For the Landau beams, the quantization of the average velocity also takes place but this effect is by three orders of magnitude less than for the LG ones. Of course, this quantization can always be calculated.

As follows from Eq. (\ref{eqphfel}), the quantization of the effective mass of the LG and Landau beams in the uniform magnetic field is given by
\begin{equation}
\begin{array}{c}
\Delta m_{eff}^{LG}\approx\frac{1}{mw_0^2},\qquad \Delta m_{eff}^{Landau}\approx\frac{1}{mw_m^2}.
\end{array}
\label{eqphfll}
\end{equation} This quantization cannot be measured directly, but it can be derived from the quantization of the average velocity.



In summary, we have revisited the theory of twisted paraxial electrons and have fulfilled their general quantum-mechanical description in the uniform magnetic field. We have generalized the Landau theory and its relativistic extensions. 
The results obtained establish fundamental properties of twisted and other structured electrons and substantially change the common view on the considered problem. Unlike all precedent investigations, we have determined structured electron states which are not plane waves along the magnetic field direction. In the weak-field limit, our solution agrees with the well-known equation for free twisted electron beams. To the contrary, the weak-field limits of the Landau wave function and its relativistic generalizations are inconsistent with this equation.
We have predicted the important observable effect of the different behavior of LG beams with the same beam waist and opposite OAM directions penetrating from the free space into the magnetic field. When $\ell_1=-\ell_2=|\ell_2|$, the final beam waists differ ($w_1'<w_2'$) and the final OAMs satisfy the relation $\ell'_1+\ell'_2>0$. Distinguishing features of the quantization of the velocity and the effective mass of the LG and Landau electrons in the uniform magnetic field have been analyzed.
A sinusoidal dependence of the beam waist on the longitudinal coordinate is the first example of spatial oscillations of the LG beams.
This effect can be compared with other unusual effects found for the LG and Gaussian beams (see, e.g., Ref. \cite{PRDqvacosc}).

\begin{acknowledgments}
This work was supported by the Belarusian Republican Foundation
for Fundamental Research
(Grant No. $\Phi$18D-002), by the National Natural Science
Foundation of China (grant No. 11975320 and No. 11805242), and by the Chinese Academy of Sciences President 's International Fellowship Initiative (grant No. 2019VMA0019).
A. J. S. also acknowledges hospitality and support by the
Institute of Modern
Physics of the Chinese Academy of Sciences. The authors are
grateful to
P. P. Fiziev for helpful discussions.
\end{acknowledgments}


\begin{thebibliography}{}

\bibitem{Bliokh2007}
K. Bliokh, Y. Bliokh, S. Savel'ev, and F. Nori,
Semiclassical Dynamics of Electron Wave Packet States
with Phase Vortices, Phys. Rev. Lett. \textbf{99}, 190404
(2007).

\bibitem{UTV}
M. Uchida, A. Tonomura, Generation of electron beams carrying
orbital angular momentum,
Nature (London) \textbf{464}, 737 (2010); 
J. Verbeeck, H. Tian, P. Schattschneider, Production and
application of electron vortex beams,
Nature \textbf{467}, 301 (2010). 

\bibitem{BliokhSOI}
K. Y. Bliokh, I. P. Ivanov, G. Guzzinati, L. Clark, R. Van
Boxem, A. B\'{e}ch\'{e}, R. Juchtmans,
M. A. Alonso, P. Schattschneider, F. Nori, and J. Verbeeck,
Theory and applications of free-electron
vortex states, Phys. Rep. \textbf{690}, 1 (2017).

\bibitem{Lloyd}
S. M. Lloyd, M. Babiker, G. Thirunavukkarasu, and J. Yuan,
Electron vortices: Beams with
orbital angular momentum, Rev. Mod. Phys. \textbf{89}, 035004
(2017).

\bibitem{LarocqueTwEl}
H. Larocque, I. Kaminer, V. Grillo, G. Leuchs, M. J. Padgett, R.
W. Boyd, M. Segev, E. Karimi,
`Twisted' electrons, Contemp. Phys. \textbf{59}, 126 (2018).

\bibitem{LloydPhysRevLett2012}
S. M. Lloyd, M. Babiker, G. Thirunavukkarasu, and J. Yuan, Electromagnetic Vortex Fields,
Spin, and Spin-Orbit Interactions in Electron Vortices, Phys. Rev. Lett. \textbf{109}, 254801 (2012).

\bibitem{Rusz}
J. Rusz, S. Bhowmick, M. Eriksson, N. Karlsson, Scattering of electron vortex beams on a
magnetic crystal: Towards atomic-resolution magnetic
measurements, Phys. Rev. B \textbf{89}, 134428 (2014).

\bibitem{Edstrom}
A. Edstr\"{o}m, A. Lubk, J. Rusz, Elastic scattering of electron vortex beams in magnetic matter,
Phys. Rev. Lett. \textbf{116}, 127203 (2016).

\bibitem{imaging}
A. B\'{e}ch\'{e}, R. Juchtmans, J. Verbeeck, Efficient creation of electron vortex beams
for high resolution STEM
imaging, Ultramicroscopy \textbf{178}, 12
(2017).

\bibitem{Observation}
V. Grillo, T. R. Harvey, F. Venturi, J. S. Pierce, R. Balboni,
F. Bouchard, G. C. Gazzadi, S. Frabboni, A. H. Tavabi, Z. Li,
R. E. Dunin-Borkowski, R. W. Boyd, B. J. McMorran, E. Karimi, Observation of nanoscale
magnetic fields using
twisted electron beams, Nat. Commun. \textbf{8}, 689 (2017).

\bibitem{OriginDemonstration}
B. J. McMorran, A. Agrawal,
P. A. Ercius, V. Grillo,
A. A. Herzing, T. R. Harvey,
M. Linck and J. S. Pierce, Origins and demonstrations
of electrons with orbital angular momentum, Phil. Trans. R. Soc. A \textbf{375}, 20150434 (2017).

\bibitem{VGRILLO}
V. Grillo, G. C. Gazzadi, E. Karimi, E. Mafakheri, R. W. Boyd, and S. Frabboni,
Highly Efficient Electron Vortex Beams Generated by Nanofabricated Phase Holograms,
Appl. Phys. Lett. \textbf{104}, 043109 (2014); V. Grillo, G. C. Gazzadi,
E. Mafakheri, S. Frabboni, E. Karimi, and R. W. Boyd, Holographic Generation of
Highly Twisted Electron Beams, Phys. Rev. Lett. \textbf{114}, 034801 (2015);
E. Mafakheri, A. H. Tavabi, P.-H. Lu, R. Balboni, F. Venturi, C. Menozzi,
G. C. Gazzadi, S. Frabboni, A. Sit, R. E. Dunin-Borkowski, E. Karimi, and V. Grillo,
Realization of electron vortices with large orbital angular momentum using miniature
holograms fabricated by electron beam lithography, Appl. Phys. Lett. \textbf{110}, 093113 (2017).

\bibitem{Bliokhmagnetic}
K. Y. Bliokh, P. Schattschneider, J. Verbeeck, F. Nori, Electron vortex beams
in a magnetic field: A new twist on Landau
levels and Aharonov-Bohm states, Phys. Rev. X \textbf{2}, 041011 (2012).

\bibitem{magnetic}
G. M. Gallatin, B. McMorran, Propagation of vortex electron wave functions in a
magnetic field, Phys. Rev. A \textbf{86}, 012701 (2012); D. Chowdhury, B. Basu, and P. Bandyopadhyay,
Electron vortex beams in a magnetic field and spin filter,
Phys. Rev. A \textbf{91}, 033812 (2015).

\bibitem{Kruining}
K. van Kruining, A. G. Hayrapetyan,
and J. B. G\"otte, Nonuniform currents and spins of relativistic electron
vortices in a magnetic field, Phys. Rev. Lett. \textbf{119}, 030401 (2017).

\bibitem{Rajabi}
A. Rajabi and J. Berakdar, Relativistic electron vortex beams in a constant
magnetic field, Phys. Rev. A \textbf{95}, 063812 (2017).


\bibitem{Greenshields}
C. Greenshields, R. L. Stamps, and S. Franke-Arnold,
Vacuum Faraday effect for electrons, New J. Phys. \textbf{14},
103040 (2012).

\bibitem{Greenshields2015}
C. Greenshields, S. Franke-Arnold, R. L. Stamps, Parallel axis theorem for free-space
electron wavefunctions, New J. Phys. \textbf{17}, 093015 (2015).

\bibitem{Landau}
L. D. Landau, Diamagnetismus der Metalle, Z. Phys. \textbf{64}, 629 (1930). 

\bibitem{LL3}
L. D. Landau, E. M. Lifshitz, \emph{Quantum Mechanics. Non-Relativistic Theory}, 3rd ed.
(Pergamon Press, Oxford, 1977), pp. 458-461.

\bibitem{Energy2a}
I. M. Ternov, V. G. Bagrov, and V. Ch. Zhukovsky, Synchrotron radiation of electron with vacuum magnetic moment, Mosc. Univ. Phys. Bull., No. 1, 21 (1966).

\bibitem{Energy2}
A. A. Sokolov and I. M. Ternov, \textit{Radiation from
relativistic electrons}, 2nd ed. (AIP, New York, 1986).

\bibitem{OConnell}
R. F. O'Connell, Motion of a relativistic electron with an anomalous magnetic moment in a constant magnetic field, Phys. Lett. A \textbf{27}, 391 (1968).

\bibitem{CanutoChiu}
V. Canuto and H. Y. Chiu, Quantum Theory of an Electron Gas in Intense Magnetic Field, Phys. Rev. \textbf{173}, 1210 (1968).

\bibitem{Energy3}
A. J. Silenko, Connection between wave functions in the Dirac and Foldy-Wouthuysen
representations, Phys. Part. Nucl. Lett. \textbf{5}, 501 (2008).

\bibitem{BBP}
S. M. Barnett, M. Babiker and M. J. Padgett, Optical orbital
angular momentum, Phil. Trans. R. Soc. A \textbf{375},
20150444 (2017).

\bibitem{Siegman}
A. E. Siegman, \emph{Lasers} (University Science Books,
Sausalito, 1986).

\bibitem{photonPRA}
A. J. Silenko, Pengming Zhang, and Liping Zou, Relativistic quantum-mechanical description of twisted paraxial electron and photon beams, Phys. Rev. A \textbf{100}, 030101(R) (2019).

\bibitem{Allen}
L. Allen, M. W. Beijersbergen, R. J. C. Spreeuw, J. P. Woerdman,
Orbital angular momentum of light and the transformation
of Laguerre-Gaussian laser modes, Phys. Rev. A \textbf{45}, 8185
(1992).

\bibitem{PlickKrenn}
W. N. Plick and M. Krenn, Physical meaning of the radial index
of Laguerre-Gauss beams, Phys. Rev. A \textbf{92}, 063841
(2015).

\bibitem{Barnett-FWQM}
S. M. Barnett, Optical Dirac equation, New J. Phys. \textbf{16},
093008 (2014).

\bibitem{Simulik} V.\,M. Simulik, I.\,Yu. Krivsky, Bosonic symmetries of the massless Dirac equation,
Adv. Appl. Clifford Alg. \textbf{8}, 
69 (1998).

\bibitem{Simulik1}
V.\,M. Simulik, I.\,Yu. Krivsky, On the extended real Clifford-Dirac algebra and new physically meaningful symmetries of the Dirac equations with nonzero mass, Reports of the National Academy of Sciences of Ukraine, No. 5, 82 (2010).

\bibitem{Simulik2}
I.\,Yu. Krivsky, V.\,M. Simulik, Fermi-Bose duality of the Dirac equation and extended
real Clifford-Dirac algebra, Condensed Matter Physics \textbf{13}, 43101 (2010).

\bibitem{Simulik3} V.\,M. Simulik, I.\,Yu. Krivsky, Bosonic symmetries of the Dirac equation,
Phys. Lett. A \textbf{375}, 
2479 (2011).

\bibitem{Simulik4}
V.\,M. Simulik, I.\,Yu. Krivsky, I.\,L. Lamer, Bosonic symmetries, solutions, and conservation laws for the Dirac equation with nonzero mass,
Ukrainian Journal of Physics 
\textbf{58}, 523 (2013).

\bibitem{Simulik5}
V.\,M. Simulik, I.\,Yu. Krivsky, I.\,L. Lamer, Application of the generalized Clifford-Dirac algebra to the proof of the Dirac equation Fermi-Bose duality,
TWMS J. Appl. Eng. Math. 
\textbf{3}, 46 (2013).

\bibitem{Simuliknew}
V.\,M. Simulik, I.\,Yu. Krivsky, I.\,L. Lamer, Some statistical aspects of the spinor field Fermi-Bose duality, Condensed Matter Physics \textbf{15}, 43101 (2012).

\bibitem{Case} K.\,M. Case, Some Generalizations of the Foldy-Wouthuysen Transformation,
Phys. Rev. \textbf{95}, 1323 
(1954).

\bibitem{Energy1}
W. Tsai, Energy eigenvalues for charged particles in a homogeneous
magnetic field -- an application of the Foldy-Wouthuysen transformation,
Phys. Rev. D \textbf{7}, 1945 (1973).

\bibitem{JMP}
A.\,J. Silenko, Foldy-Wouthuysen transformation for
relativistic particles in external fields, J. Math. Phys. {\bf 44}, 2952 (2003).

\bibitem{JINRLett12}
A. J. Silenko, Classical limit of relativistic quantum mechanical equations in the Foldy-Wouthuysen representation, Pis'ma Zh. Fiz. Elem. Chast. Atom. Yadra \textbf{10},
144 (2013) [Phys. Part. Nucl. Lett. \textbf{10}, 91 (2013)].

\bibitem{Reply2019}
A. J. Silenko, Pengming Zhang, and Liping Zou, Silenko, Zhang, and Zou Reply, Phys. Rev. Lett. \textbf{122}, 159302 (2019).

\bibitem{PRAFW}
Liping Zou, Pengming Zhang, and A. J. Silenko, Position and spin in relativistic quantum mechanics, Phys. Rev. A \textbf{101}, 032117 (2020).

\bibitem{paraxialLandau}
Liping Zou, Pengming Zhang, and A. J. Silenko, Paraxial wave function and Gouy phase for a relativistic electron in a uniform magnetic field, J. Phys. G: Nucl. Part. Phys. \textbf{47}, 055003 (2020).

\bibitem{Kogelnik}
H. Kogelnik and T. Li, Laser Beams and Resonators, Appl. Opt. \textbf{5}, 1550 (1966).

\bibitem{Alda}
J. Alda, Laser and Gaussian Beam Propagation and Transformation, in \emph{Encyclopedia of Optical Engineering}, vol. 2, ed. by R. G. Driggers, C. Hoffman, and R. Driggers (Marcel Dekker Inc., New York, 2003), pp. 999-1013.

\bibitem{Pampaloni}
F. Pampaloni, J. Enderlein, Gaussian, Hermite-Gaussian, and
Laguerre-Gaussian beams: A primer, arXiv:physics/0410021 (2004).

\bibitem{SupplementalMaterial}
See Supplemental Material at

\bibitem{GradshteynRyzhik}
I. S. Gradshteyn and I. M. Ryzhik, \emph{Table of
Integrals, Series, and Products}, 8th ed. (Academic Press, Amsterdam, 2015).

\bibitem{Loffler}
S. L\"{o}ffler, A.-L. Hamon, D. Aubry,
P. Schattschneider, A quantum propagator for electrons in a round magnetic lens, Adv. Imag. Elect. Phys. \textbf{215}, 89 
(2020).

\bibitem{FloettmannKarlovets}
K. Floettmann and D. Karlovets, Quantum mechanical formulation of the Busch theorem, Phys. Rev. A \textbf{102}, 043517 (2020).

\bibitem{PhysRevLettLanzhou2019}
A. J. Silenko, Pengming Zhang, and Liping Zou, Electric Quadrupole Moment and the Tensor Magnetic Polarizability
of Twisted Electrons and a Potential for their Measurements, Phys. Rev. Lett. \textbf{122}, 063201 (2019).

\bibitem{Barut}
A. O. Barut and A. J. Bracken, Magnetic-moment operator of the relativistic electron, Phys. Rev.
D \textbf{24}, 3333 (1981).

\bibitem{ResonanceTwistedElectrons}
A. J. Silenko, Pengming Zhang and Liping Zou, Relativistic quantum dynamics of twisted electron beams in arbitrary electric and magnetic fields, Phys. Rev. Lett. \textbf{121}, 043202 (2018).

\bibitem{NPCS2018}
A. J. Silenko, Pengming Zhang and Liping Zou, Classical and Quantum-Mechanical Description of
Dynamics of Twisted Electrons in Electromagnetic Fields, Nonlinear Phenom. Complex Syst. \textbf{21}, 371 (2018).

\bibitem{Schattschneider}
P. Schattschneider, Th. Schachinger, M. St\"{o}ger-Pollach, S. L\"{o}ffler, A. Steiger-Thirsfeld,
K.Y. Bliokh and  F. Nori, Imaging the dynamics of free-electron
Landau states, Nature Commun. {\textbf5}, 4586 (2014).

\bibitem{Karlovets}
D. Karlovets, Relativistic vortex electrons: Paraxial versus nonparaxial regimes, Phys. Rev.
A \textbf{98}, 012137 (2018).

\bibitem{OAMdiameter}
K. Saitoh, Y. Hasegawa, N. Tanaka and M. Uchida, Production of electron vortex beams carrying large orbital
angular momentum using spiral zone plates,  J. Electron Microsc. \textbf{61}, 
171 (2012).

\bibitem{PRDqvacosc}
F. Karbstein and E. A. Mosman, Enhancing quantum vacuum signatures with tailored laser beams, Phys. Rev. D \textbf{101}, 113002 (2020).

\end{thebibliography}
\end{document}